\documentclass[manuscript]{aastex}

\newcommand{\HST}{{\it HST}}
\newcommand{\Gaia}{{\it Gaia}}

\begin{document}          


\title{The Orbit of the Close Companion of Polaris:  {\em Hubble Space
Telescope\/} Imaging 2007 to 2014\altaffilmark{*} \\
\quad \\
}

\author{
Nancy Remage Evans\altaffilmark{1},
Margarita Karovska\altaffilmark{1},  
Howard E. Bond\altaffilmark{2,3},
Gail H. Schaefer\altaffilmark{4}, 
Kailash C. Sahu\altaffilmark{3},
Jennifer Mack\altaffilmark{3},
Edmund P. Nelan\altaffilmark{3},
Alexandre Gallenne\altaffilmark{5},
and
Evan D. Tingle\altaffilmark{1} 
}  

\altaffiltext{*}
{Based on observations made with the NASA/ESA {\it Hubble Space Telescope},
obtained by the Space Telescope Science Institute. STScI is operated by the
Association of Universities for Research in Astronomy, Inc., under NASA contract
NAS5-26555.}

\altaffiltext{1}
{Smithsonian Astrophysical Observatory, MS 4, 60 Garden St., Cambridge, MA
02138, USA; nevans@cfa.harvard.edu}

\altaffiltext{2}
{Department of Astronomy \& Astrophysics, Pennsylvania State University,
University Park, PA 16802, USA; heb11@psu.edu}

\altaffiltext{3}
{Space Telescope Science Institute, 3700 San Martin Drive, Baltimore, MD
21218, USA}

\altaffiltext{4}
{The CHARA Array of Georgia State University, Mount Wilson Observatory,
Mount Wilson, CA 91023, USA}

\altaffiltext{5}
{European Southern Observatory, Alonso de Cordova 3107, Casilla 19001, 
Santiago, Chile}




\begin{abstract}

As part of a program to determine dynamical masses of Cepheids, we have imaged
the nearest and brightest Cepheid, Polaris, with the {\it Hubble Space
Telescope\/} Wide Field Planetary Camera~2 and Wide Field Camera~3. Observations
were obtained at three epochs between 2007 and 2014. In these images, as in
\HST\/ frames obtained in 2005 and 2006 which we discussed in a 2008 paper, we
resolve the close companion Polaris~Ab from the Cepheid Polaris~Aa. Because of
the small separation and large magnitude difference between Polaris Aa and Ab,
we used PSF deconvolution techniques to carry out astrometry of the binary.
Based on these new measurements, we have updated the elements for the 29.59~yr
orbit. Adopting the distance to the system from the recent {\it Gaia\/} Data
Release~2, we find a dynamical mass for the Cepheid of $3.45 \pm 0.75\,M_\odot$,
although this is preliminary, and will be improved by  CHARA 
measurements covering periastron. 
As is the case for the recently determined dynamical mass for the Cepheid
V1334~Cyg, the mass of Polaris is significantly lower than the ``evolutionary mass''
predicted by fitting to evolutionary tracks in the HR diagram. We discuss
several questions and implications raised by these measurements, including the
pulsation mode, which instability-strip crossing the stars are in, and possible
complications such as rotation, mass loss, and binary mergers. The distant third
star in the system, Polaris~B, appears to be older than the Cepheid, based on
isochrone fitting. This may indicate that the Cepheid Polaris is relatively old
and is the result of a binary merger, rather than being a young single star. 

 \end{abstract}




\section{Introduction}

The link between Cepheids and the extragalactic distance scale illustrates the
importance of a thorough understanding of Cepheid physics. In addition to their
significance as standard candles, they are benchmarks for confronting
stellar-evolution theory. They ultimately
become white dwarfs, which means a subset will be compact objects in
binary\slash multiple systems, a part of a population from which exotic objects
are formed (novae, supernovae, X-ray binaries). The long-standing ``Cepheid mass
problem'' (the mass predictions from evolutionary tracks are larger than 
predictions from pulsation calculations) has been reduced 
through revision of interior opacities (e.g., Bono et al.\ 2001), but still
persists. Measured dynamical masses for Cepheids, coupled with a well-determined
luminosity, are needed to confront these questions as well as to provide
predictions about the origin of late-stage exotic objects. 




Polaris ($\alpha$~Ursae Minoris) is the nearest and brightest Cepheid. It is a
member of a triple system, with its resolved 8th-mag F3~V physical companion,
Polaris~B, lying at a separation of $18''$. The Cepheid itself has been known
for many years to be a single-lined spectroscopic binary with a period of about
30~years (Roemer 1965; Kamper 1996, hereafter K96, and references therein),
whose components we denote Polaris~Aa and Ab. Thus Polaris offers an opportunity
for direct determination of a Cepheid's dynamical mass. To date, V1334\,Cyg is
the only other Cepheid with a measured
purely dynamical mass (Gallenne et al.\ 2013;  2018) in the Milky Way;
however interferometric studies which  resolve orbits will increase the
number over the next few years (Gallenne et al.\ 2015).  
Cepheids in eclipsing binaries in 
 the Large Magellanic Cloud summarized by Pilecki et al. (2018)
provide an important comparison.


Recent papers have disagreed about the distance of the Polaris system. The {\it
Hipparcos\/} mission (van Leeuwen 2007, hereafter vL07) measured an absolute
parallax of $7.54\pm0.11$~mas ($d = 132.6\pm1.9$~pc). However, Turner et al.\
(2013, hereafter T13, and references therein) gave astrophysical arguments that
the parallax of Polaris must be considerably larger, $10.10\pm0.20$~mas
($d=99\pm2$~pc), but this was disputed by van Leeuwen (2013). 
More recently, Bond et al.\ (2018, hereafter B18) measured a
parallax of only $6.26 \pm 0.24$~mas ($d=160 \pm 6$~pc), using the Fine Guidance
Sensors (FGS) on the {\it Hubble Space Telescope\/} (\HST)\null. Since Polaris
itself is too bright for FGS measurements, B18 determined the parallax of the
wide physical companion, Polaris~B\null. The large distance based upon the FGS
parallax implies a relatively high luminosity, indicating that Polaris pulsates
in the second overtone rather than the fundamental mode. For the {\it
Hipparcos\/} parallax, the pulsation is likely to be at the first overtone
(e.g., Feast \& Catchpole 1997; Bono et al.\ 2001; van Leeuwen et al.\ 2007;
Neilson 2014). For the large parallax advocated by T13, the pulsation would be
in the fundamental mode, as they argued. 


In the recent Data Release~2 (DR2), 
the {\it Gaia\/} mission derived a parallax
of $7.292 \pm 0.028$~mas  for Polaris~B (Gaia Collaboration et al.\ 2016,
2018a, 2018b).
Polaris itself is too bright to have been measured for DR2. This result  
corresponds to a distance of  137.2 pc with a  range of 136.61 to 137.67 pc.
Applying the correction found by Lindegren et al.\ (2018) of +0.03~mas to the
parallax reduces the distance to 136.6 pc.  Polaris B has a magnitude $V =
8.65$, well below  the brightness limit for {\it Gaia}.  The discussion by Riess
et al.\ (2018)  of Cepheids finds a slightly larger zeropoint offset
($-0.046$~mas) for the Cepheids they discuss,  although Polaris~B  is somewhat
hotter (F3~V) and the zeropoint  may have color dependence. Further discussion 
is given in Section 3.


Polaris is very unusual among Cepheids, in that its pulsation amplitude
decreased over much of the 20th century. However, in recent years the amplitude
has begun to increase again (e.g., Bruntt et al.\ 2008). This may indicate a
very long-period pulsation\slash interference cycle. 


Our team obtained near-ultraviolet images of Polaris with the  High-Resolution
Channel (HRC) of the Advanced Camera for Surveys (ACS) onboard \HST\/ in 2005
and 2006. These images succeeded in resolving the Polaris Aa and Ab pair for the
first time, at a magnitude difference of 5.4~mag and a separation of about
$0\farcs17$ (Evans et al.\ 2008, hereafter E08). This allowed us to estimate the
first purely dynamical mass for any Cepheid (albeit with a large uncertainty).
Since then, we have used \HST\/ cameras to make three more observations, which
now cover a larger portion of the orbit. However, the new measurements were
considerably more difficult and inexact, because the separation between Polaris
and its close companion has been decreasing, and because the HRC was no longer
available. In this paper, we report these new measurements, and update the
dynamical-mass estimates. 


\section{Observations, Image Processing, and Deconvolution}

Table 1 lists information about our three new \HST\/ observations, along with
the previous 2005 and 2006 measurements given by E08. The new \HST\/
observations reported here were made in 2007, 2009, and 2014. In this section we
give details of our observing strategy, the creation of combined images at
each epoch, and finally the deconvolution of the images and astrometry of the
binary.


\subsection{Observing Strategies}

Two different \HST\/ cameras were used for our new observations, as follows:

(1)~{\it WFPC2 2007}. Following the successful detection by E08 of the close
companion of Polaris with ACS/HRC in 2005 and 2006, we had planned to obtain an
additional HRC observation in 2007. Unfortunately, the ACS suffered a failure in
2007 January, making it unavailable. We used instead the Wide Field Planetary
Camera~2 (WFPC2), and the observations were made on 2007 July~17. We chose the
near-ultraviolet broad-band F218W filter, both because Polaris~Ab is slightly
hotter than the Cepheid, and because of the smaller point-spread function (PSF)
at shorter wavelengths. We used a similar observing strategy as for the HRC
observations, placing the target in the Planetary Camera (PC) CCD\null. The PC
plate scale is $0\farcs046\,\rm pixel^{-1}$, as compared with the ACS/HRC scale
of $0\farcs026\,\rm pixel^{-1}$. We first obtained a series of eight short
(0.8~s) exposures spread over a three-point dither pattern. The telescope
orientation was specified such that Polaris~A and B would lie parallel to the
edge of the PC field and symmetrically placed relative to the corners, and we
positioned these targets close to the edge so as to minimize
charge-transfer-inefficiency effects. This orientation also placed the location
of Ab away from the diffraction spikes of Aa, for its anticipated position angle
(P.A.)\null. We then moved the telescope to put the wide companion Polaris B at
the same position used for A, and obtained six longer exposures (50~s) in the
same dither pattern. This provided a PSF standard at the same field position as
Polaris~A, exposed to about the same count level.


(2)~{\it WFC3 2009}. The WFPC2 camera was removed from \HST\/ during the 2009
May Servicing Mission~4, and replaced with the Wide Field Camera~3 (WFC3). We
observed Polaris with the WFC3 UVIS channel on 2009 November~18. The UVIS plate
scale is $0\farcs040\,\rm pixel^{-1}$.  Because of the increased sensitivity of
WFC3 relative to WFPC2, we used the narrow-band near-ultraviolet filter FQ232N,
in order to avoid saturating the image of Polaris~Aa. We again chose a telescope
orientation to place Ab between the diffraction spikes of Aa. We obtained a
total of nine dithered exposures of 0.5 and 0.7~s, with Polaris~A placed near
the center of a $2048\times2048$-pixel subarray ($81''\times81''$). Then we
placed Polaris~B near the center of the field, and obtained five dithered
exposures of 45~s each to use as a PSF reference.


(3)~{\it WFC3 2014}. By the time of our final \HST\/ observations of Polaris, it
had become known that short exposures with WFC3 are affected by vibrations in
the camera due to the shutter mechanism (``shutter jitter''; see Sahu et al.\
2014 for a technical discussion). Exposures with the shutter in the ``A'' blade
position are less affected than those in the ``B'' position, so we used the
newly available {\tt BLADE=A} option for all of our exposures.  Instead of using
longer exposures on Polaris~B as a PSF reference star, which would be much less
affected by shutter jitter, we observed $\gamma$~Cygni---a star with nearly the
same magnitude, color, and spectral type as the Cepheid---immediately after the
Polaris observation. We obtained 14 dithered FQ232N exposures on Polaris of
1.5~s each (which the 2009 exposures indicated would not be saturated), and
chose a telescope orientation that placed Polaris~Ab at an optimum location in
the PSF to avoid diffraction spikes and filter ghosts, based on our analysis of
the 2009 images. This was followed by a set of 12 essentially identical dithered
exposures on $\gamma$~Cyg. The observations were otherwise very similar to those
made in 2009.

Our first attempt at these observations (2014 March~17) suffered a loss of
guide-star lock partway through the $\gamma$~Cyg visit, so that we did not have
useful PSF images. These data were not included in our data analysis. Both sets
of observations were repeated successfully on 2014 June~26. 




\subsection{Combining the Dithered Images}





As described above, we obtained a set of dithered exposures of Polaris~A and B
at two epochs, and of Polaris~A and $\gamma$~Cyg at a single epoch. In order to
combine these images into optimal single images at each epoch, we used the {\tt
AstroDrizzle} package developed at the Space Telescope Science Institute
(STScI), running under {\tt PyRAF}\footnote{{\tt AstroDrizzle} and {\tt PyRAF}
are products of the Space Telescope Science Institute, which is operated by AURA
for NASA.}.  The resulting combined images have a pixel scale of $0\farcs02$. 
For the WFC3 data in 2009 the problem of ``shutter jitter'' was a concern.  We 
experimented  with the images, and found that the ``even'' images in 
the dither pattern were better quality, and were able to  test  the effect 
on the location of the companion.

%
%
%
%
%
%
%
%

\subsection{Deconvolution and Astrometry of Polaris~Ab}

The observed stellar images are blurred by the PSF of the telescope and camera
optics, and the finite size of the detector pixels. To remove the effects of
this blurring we have in principle several options, including a subtraction of
the PSF (based on images of a reference point source), fitting of a model for
the source plus PSF to the observed source, or a full deconvolution of the PSF.

Because of the small separation, the large magnitude difference between the
Cepheid Polaris~Aa and its companion Polaris~Ab, and the larger plate scales
compared to the ACS/HRC used in our earlier observations, we could not use
direct PSF model fitting for any of the more recent observations. Direct PSF
subtraction was also unable to detect the companion. Therefore, we applied
deconvolution/restoration 
techniques, using the observed single point-like reference stars
as models of the PSF.

We applied two image-restoration methods. One of them is the well-known
Richardson-Lucy (R-L) deconvolution technique (Richardson 1972; Lucy 1974), and
the other is a statistical deconvolution (restoration)
technique called EMC2 (Expectation
through Markov Chain Monte Carlo).
 The EMC2 technique uses a wavelet-like multiscale
representation of the true image in order to achieve smoothing at all scales of
resolution simultaneously (see Esch et al.\ 2004; Karovska et al.\ 2005, 2010).

As described above, for the 2007 and 2009 observations we used Polaris~B as a
PSF reference star, which we had placed at the same field location as
Polaris~A\null. For these two epochs, the R-L technique successfully resolved
the companion. Figures~1 and 2 depict the results of the deconvolutions, with
the locations of the companion Polaris Ab marked with circles, connected by
arrows to the center of Polaris~Aa. The last two columns in Table~1 give the
derived P.A. and separation of Polaris~Ab. The quoted errors estimated by eye
correspond to one pixel in the drizzled image ($0\farcs02$).  This is conservative
 for the 2009 WFC3 observation, where they may be smaller.    



By the time of the 2014 observation, Polaris~Ab had moved much closer to Polaris
Aa, making it considerably more difficult to separate the two components. We
did, however, have images of the PSF reference star $\gamma$~Cyg, obtained with
the same exposure times as Polaris, and thus affected by the same amount of
shutter jitter. For this observation, we applied the EMC2 deconvolution
technique, which is better suited for detecting faint companions.

Given the small separation and large magnitude difference, we had to search for
Polaris~Ab amidst the PSF structure and noisy remnants of the deconvolution.
However, our search was guided by a predicted location of the companion, based
on an existing orbital fit to the previous measurements. This facilitated the
search for the companion, and we believe we have detected it.

In Figure 3 we show the deconvolved 2014 image. A circle marks the region where
we believe we may have detected the companion. Also included are two smaller
circles indicating the predicted locations.  The first uses the 
orbit from the spectroscopic orbit, our \HST\/ observations from 2005 to 2007,
and the astrometric measurements from Wielen et al.\ (2000).  
The second uses a slightly modified orbit with 1$\sigma$ change in the
inclination. The resulting P.A. and separation are listed in the bottom row in
Table~1. The quoted error corresponds to the drizzled pixel size of $0\farcs02$.
Although the 2014 detection is not as clear as in the earlier observations, it
may provide an additional constraint on the orbit. Unfortunately, for the
remaining lifetime of \HST, the separation will be too close to be resolved.


\section{Visual Orbit for Polaris Aa--Ab}

Our additional \HST\/ measurements of Polaris Aa--Ab substantially improve the
orbital coverage, compared with the results presented by E08. We have fitted 
new orbits to the five astrometric data points presented in Table~1, and also to
a subset of only the first four points, using a Newton-Raphson method to
minimize $\chi^2$ by calculating a first-order Taylor expansion for the
equations of orbital motion. Since we have only four or five astrometric
observations, covering only a small portion of the orbit, we fixed the values of
the spectroscopic parameters of the orbital period $P$, eccentricity $e$, and
longitude of periastron passage $\omega$, to those determined by K96. The date
of periastron passage $T_0$ reported by K96 was discussed and updated by Wielen
et al.\ (2000), and we adopt their value. We then solved for the remaining
parameters of the angular semi-major axis $a$, orbital inclination $i$, and the
position angle of the line of nodes~$\Omega$. We obtained formal errors for each
of the free parameters from the diagonal elements of the covariance matrix. To
account for the uncertainties in the spectroscopic orbital parameters, we
computed a series of 10,000 orbital solutions where we randomly varied $P$,
$T_0$, $e$, and $\omega$ within their 3$\sigma$ uncertainties, and solved for
the best-fit values of $a$, $i$, and $\Omega$ for each iteration. We computed
the standard deviation for each parameter's distribution, and added these
uncertainties in quadrature to the formal errors computed from the best-fit
solution when fixing the spectroscopic parameters. 

We present two orbital solutions. One uses all of the \HST\/ data, and the other
omits the 2014 measurement, which is the most uncertain. The final orbital
parameters and uncertainties are presented in Table~2 for both solutions.

In Figure~\ref{fig.orbit} we show the \HST\/ measurements compared with the
best-fit orbit for all data.  
The shared gray area shows the range of orbital solutions that
fit the data. This region was computed using two different methods.  As
described above, in the first method we randomly varied the spectroscopic
parameters within their 1$\sigma$ uncertainties, and re-derived the best-fitting
orbit for each iteration. In the second method, we fixed the spectroscopic
parameters and randomly varied $a$, $i$, and $\Omega$ in order to search for
orbital solutions within $\Delta\chi^2 = 3.53$ of the minimum, corresponding to
a 68.3\% confidence interval, for the three free parameters. The shaded region
in the figure represents the maximum deviations in orbital solutions obtained
from these two samples.  The residuals in separation and P.A. compared with the
best-fit orbit are plotted in Figure~\ref{fig.residual}.


By assuming a parallax $\pi$ for the system, we can then compute the total mass
of Polaris Aa and Ab through Kepler's Third Law: $M_{\rm tot} = M_{\rm Aa} +
M_{\rm Ab} = a^3/(\pi^3 \, P^2)$, where the masses are in solar units,  $a$ and
$\pi$ in seconds of arc, and $P$ in years. Using the radial-velocity amplitude
of Polaris~Aa ($K_{Aa} = 3.72 \pm 0.03\,\rm km\,s^{-1}$) from the spectroscopic
orbit of K96, we can derive individual masses of Aa and Ab. 
We show the
resulting dynamical masses using both the {\it Gaia\/} parallax and the 
{\it Hipparcos\/} parallax in Table~3. 
We note that the errors for the distance from {\it Gaia} 
 are small, but the mass itself depends on the orbital solution, which 
 will become  better defined during the periastron passage.  The comparison 
in Table 2 of the orbit omitting the least accurate point illustrates 
the possible difference in the solutions.  

The mass derived depends heavily on the parallax. 
The parallax from {\it Gaia} DR2 for Polaris B 
has reasonable ``excess noise" of 0.0 and a reduced $\chi^2$ of
2.7.  The ``goodness of fit" is 12.2, which is higher than
the desirable 9, but within a reasonable range.  
The use of 
a zero point correction to {\it Gaia} parallaxes adds 
to the uncertainty.  
The mass of the Cepheid without the correction becomes 
3.50$\pm$0.76\,M$_\odot$.   We anticipate improvement
in the mass determination from subsequent {\it Gaia} releases
such as  DR3, as 
well as more complete observations of the orbit.


We note two further points about the adopted parallax for the Polaris system.
(1)~Using the T13 parallax would result in a mass  for the Cepheid of only
$1.08\,M_\odot$, far smaller than either the observational or theoretical
predictions for a Cepheid. We believe the large parallax advocated by T13 is now
definitively ruled out. (2)~We recomputed the FGS parallax reported by B18, by 
adopting as input values the parallaxes and proper motions for the background
reference stars as given in the {\it Gaia\/} DR2. This actually makes the
derived parallax of Polaris~B even smaller than given by B18. At this point,
given the good agreement between the {\it Hipparcos\/} parallax for Polaris~A
and the DR2 parallax for Polaris~B, we no longer advocate the FGS parallax. We
are examining reasons why it  might be incorrect (such as the possibility that
Polaris~B could be a marginally resolved close binary).

Table~3 also lists, in the bottom line, the mean visual absolute magnitude of the
Cepheid Polaris~Aa, calculated by assuming a mean apparent magnitude of 
$\langle V\rangle = 1.982$ and a reddening of $E(B-V)=0.01\pm0.01$ (see B18 and
references therein), and adopting the two different parallaxes.

%
%
%


\section{Discussion}

\subsection{ Properties}





Based on the updated results given in this paper
(which still covers only a fraction of the entire 29.95~yr orbit), we now
compare the 
masses in Table~3 with expectations from current Cepheid calibrations.
For the Cepheid, Polaris Aa, the mass prediction is compared below 
 with the measured mass of the Cepheid V1334 Cyg (Gallenne et al.\ 2018)   
as well as with evolutionary tracks.
  There is also
information about the astrometric companion, Polaris Ab, which was 
resolved by the \HST\/ HRC of the ACS at an epoch  when the separation was 
wider.  In E08, its mass was estimated based on photometric measurements
to be $1.3\,  M_\odot$.  
This is in reasonable agreement with the estimate from the orbit (Table~3)
but more precise masses for the Cepheid and the companion await more 
complete coverage of the
orbit.
The properties of both the Ab and B components of the system are
discussed further compared with 
an isochrone in the discussion of mergers below.  

\subsection{Questions and Implications}

As the nearest Cepheid, Polaris has  measured properties which are available for 
only a few other Cepheids which provide a number of distinctive 
clues  and raise questions which may be answered as the orbit
becomes better defined.  
In this section we discuss  recent results in  several areas
 relevant to the interpretation of a measurement of the mass (and luminosity) of
Polaris.   The most important comparison with the preliminary 
mass of Polaris is with the recently determined mass of V1334 Cyg, and a 
comparison of these two stars with evolutionary tracks.  Subsequently we discuss the 
period change (often interpreted as a direct 
measure of evolution),
instability strip crossing, possible mass loss or  a previous merger, 
and implications from abundances.

\subsubsection{Pulsation Mode}

Fig.~\ref{pol.mode} shows the location of Polaris in the Leavitt
period-luminosity relation,  using the  {\it Gaia\/} absolute magnitude $M_V$.  
The comparison Cepheids have  FGS trigonometric parallaxes (Benedict et al.\
2007), \HST\/ spatial-scan  parallaxes (Riess et al.\ 2018b), and the
light-echo distance of RS Pup (Kervella et al.\ 2014).  $\langle V\rangle$ and 
$E(B-V)$ have
been taken from the Galactic Cepheid  Database (Fernie et al.\ 1995). While  
{\it Gaia\/} results for a large Cepheid sample  will undoubtedly contribute
immensely to the Leavitt Law, the brightest Cepheids are challenging for  {\it
Gaia},  so we use the previous \HST\/ results  for this figure. The  result
from   {\it Gaia\/} for Polaris B and hence the  Cepheid Polaris Aa is  consistent
with pulsation in the first-overtone  mode (though second overtone is still
within the errors).  We will treat  the Cepheid as a first-overtone pulsator
below.

\subsubsection{Comparison with V1334 Cyg}

V1334 Cyg is  another overtone pulsator with a  similar (fundamentalized)  period 
(4.74$^d$)  and luminosity
to Polaris.  Its  mass and distance have been measured by Gallenne et al.\ (2018)
to unprecedented accuracy using 
interferometry combined with radial velocities from the visible and ultraviolet, 
 based on the orbit of Evans (2000).  
They find a Cepheid mass of $4.288 \pm 0.133 \,M_\odot$ and a distance of 
$720.35 \pm 7.84$~pc, corresponding to an absolute magnitude $M_V = -3.37$
(including a correction for the companion),
which probably provides the best estimate of the expectation for Polaris. 
 Polaris has a somewhat longer period than V1334 Cyg
and a slightly brighter $M_V = -3.73 \pm 0.03$.  The mass of the Cepheid Aa
({\it Gaia}; Table 3) is smaller ($3.45 \pm 0.75 \,M_\odot$), but 
the  mass will not be final until the orbit is more fully covered.     
The mass of Polaris Aa only differs by a little more than 1$\sigma$ from that of V1334 Cyg
and the luminosities are similar. 
We discuss the rapid period change and variation in pulsation amplitude of 
Polaris below but 
we draw attention to the fact that V1334 Cyg shows no indication of period 
change, at least since 1968 (Berdnikov et al.\ 1997). 
In the previous discussion (B18), when Polaris appeared to be a second-overtone 
pulsator, it seemed likely that  the different pulsation mode 
might be responsible for the difference in the envelope pulsation
properties (period change and amplitude variation).  However it now seems
most likely that both stars are pulsating in the same mode, 
illustrating a range of envelope pulsation properties.   


\subsubsection{Evolutionary Tracks} 

The combination of observed mass and luminosity is needed for
 comparison with  evolutionary tracks. 
For a number of years it has been clear that  a 
moderate amount of core convective overshoot in intermediate-mass
main-sequence stars increases the fuel available and ultimately the 
luminosity of the ``blue loops'', the stage at which Cepheids are found.  (See 
Nielson et al.\ 2012 for examples of this effect.)  This has the 
unfortunate counter effect of decreasing the temperature range of the blue
loops  so that typically evolutionary tracks for 
stars less massive than 5 $M_\odot$ do not penetrate the 
instability strip (e.g., Fig.~2 in Neilson et al.\ 2012a).  Recently 
an additional parameter has been explored in detail:
 rotational velocity.  Anderson et al.\ (2014) provide  
evolutionary tracks for a range of initial rotations, which produce 
results similar to convective overshoot calculations: more luminous 
blue loops for a given mass.  However, there is  less suppression of 
loops for lower mass stars.
Either of these possibilities or a combination
may ultimately prove to be an accurate 
description, and the blue loops are notoriously sensitive to 
input parameters.  However, at present it is not clear whether either 
theoretical approach matches the masses available or those expected
in the near future.  

In Fig.~\ref{ev.pol} we show the comparison between the evolutionary tracks with 
rotation discussed in Anderson et al.\ (2014) which are taken from 
Georgy et al.\ (2013).  Polaris and V1334 Cyg have luminosities from 
{\it Gaia} (Table 3) and Gallenne et al.\ (2018) respectively.  Temperatures
are from $(B-V)_0$ (Evans 1988 and Evans et al.\ 2013 respectively) and the 
temperature color calibration as discussed in Evans \& Teays (1996). 
 However, 
 masses inferred from the luminosities of the evolutionary tracks 
in Fig.~\ref{ev.pol} are about $5\,M_\odot$,  which is larger than the 
measured values, even for V1334 Cyg.  This is true even when the stars start 
their main-sequence life with a large rotational velocity.  
The mass of Polaris estimated using the evolutionary tracks in 
 Fig.~\ref{ev.pol} (the ``evolutionary mass'') 
 is approximately $5.5\, M_\odot$ for the tracks
incorporating maximum rotation, but $6.0\, M_\odot$ without rotation.
These values are discrepant by 2.7$\sigma$ and 3.4$\sigma$ from the 
measured mass of 3.45 +/- 0.75 Msun (Table 3).  However, only a small 
portion of the visual orbit is currently covered; a systematic offset  
of 3$\sigma$ in the semi-major axis as the periastron passage is mapped 
in future observations would bring the mass of Polaris in agreement with 
the evolutionary masses.

In Fig.~\ref{ev.pol.blu} we  compare several representative evolutionary tracks
in the ``blue loop'' section of the HR diagram.  Specifically, 5 $M_\odot$
tracks are plotted from  the Geneva series (Georgy et al.\ 2013), the MIST
series (Choi et al.\ 2016) and  the PARSEC set (from \url{{\tt
https://philrosenfield.github.io/padova\_tracks}}); based on the results from
the Bressan et al.\ (2012) group.  We selected  5 $M_\odot$ tracks because they
cover the temperature range of the instability strip and are available for a
range of rotational  velocities from several codes.  For all three codes, tracks
are shown with zero  initial rotation.  For   Geneva and MIST, a track with
substantial initial rotation is also shown (0.95 critical  breakup speed for
Geneva,  0.4 critical velocity for MIST).  Fig.~\ref{ev.pol.blu}  shows the
resulting variation  in luminosity from  the different codes as well as
differing predictions for the incorporation of main sequence rotation.  

V1334 Cyg is the first Cepheid with a dynamically determined mass with  a small
enough uncertainty ($4.288 \pm0.133\,M_\odot$) to strongly constrain the tracks. 
The most important point in Fig.~\ref{ev.pol.blu} is that V1334 Cyg is  
{\it significantly\/} more  luminous than the predictions for its mass 
from the calculations
regardless of the  code or rotation.  An alternate description is that the
Geneva tracks in Fig.~\ref{ev.pol} indicate that its mass is approximately 5
$M_\odot$, which is 0.7 $M_\odot$ larger than the measured mass.  

A further note about the mass of V1334 Cyg is that   
it strongly constrains the lower mass at which a star destined to become a 
Cepheid can enter the instability region after the red giant branch.  
Current tracks in
Fig.~\ref{ev.pol} show that a 4 $M_\odot$ star would 
not become hot enough.  This is doubly true in that overtone pulsators 
are found among shorter period stars (even when ``fundamentalized'')
on the blue side of the instability strip.  Thus, a track has to become 
hot enough to reach the blue  region of the Cepheid instability strip.  


Another parameter which is directly comparable with the predictions of
evolutionary tracks is the radius.  M\'erand et al.\ (2006) measured an 
angular diameter of Polaris of 3.123 $\pm$ 0.008 mas,  
which corresponds to 46  $R_\odot$ using the {\it Gaia\/} distance.
We can make an approximate comparison between this and the results of 
the evolutionary tracks (Anderson et al.\ 2016) using the 
luminosity and temperature  tabulated in their Table A.4. In the
comparison we use the Milky Way metallicity, an initial rotation 
which is half critical breakup rotation, a mass of $5\,M_\odot$,
and first-overtone pulsation. 
The 5$M_\odot$ evolutionary track 
does not cover the full temperature width of the instability strip for 
the second and third crossing (discussed below), but rather enters the cool edge and then 
turns in the middle and returns to the cool edge, being the lowest mass of 
the computed tracks to enter the instability strip.  However, the truncated
second and third crossings cover the range of radii of 42 to 52 $R_\odot$
while the first crossing  spans 26 to 33  $R_\odot$.  Thus,  in this
approximate comparison, the third crossing (increasing period)  is 
in reasonable agreement with the measured radius.  

\subsubsection{Period Change}  

Polaris undoubtedly has a changing period
(Neilson et al.\ 2012b and references therein).
  A changing period is typically interpreted as resulting 
from evolution through the instability strip; however
there are a number of indications that more than this simple interpretation is 
needed.   
Overtone pulsators appear to have higher rates 
of period change than fundamental-mode pulsators of comparable period 
(Szabados 1983; Evans et al.\ 2002).  
This would, in fact, explain what is described as the anomaly of large period 
changes in short-period Cepheids by Turner et al.\ (2006). 
Studies using spectacular satellite observations ({\it Kepler, CoRot, MOST, WIRE}) are
alerting us to  hitherto unknown pulsation behavior: new modes
 and interactions. 
For instance, first-overtone Cepheids in the Small Magellanic Cloud have peculiarities 
which may be related to  nonradial pulsation (Smolec \& Sniegowska 2016). 
In addition, continuous sequences of high-quality 
photometry with the {\it MOST\/} satellite of RT Aur (fundamental mode) and 
SZ Tau (first overtone) confirm that overtone cycles are less stable than 
fundamental mode cycles (Evans et al.\  2015), though this has been 
disputed by Poretti et al.\ (2015).

Thus, while Polaris's rapid period change has sometimes been interpreted as
indicating a  first rather than a third crossing of the instability strip
(Turner et al.\ 2005), the working model here is that for overtone pulsators,
period changes are not {\it strictly\/} due to evolution through the instability
strip.  Rather period changes are at least partially due to some aspect of
pulsation that we do not yet fully understand.  Other characteristics of 
Polaris's pulsation  support this.  While the period change O-C (observed minus
computed) diagrams suggest a parabola, the actual form is more complex, with a
``glitch'' about 40 years ago (Turner et al.\ 2005; Neilson et al.\ 2012).  This
requires  two tightly constrained parabolas to fit  the data.  Furthermore, the
pulsation amplitude  of Polaris has a marked drop followed by a gradual increase
over many decades,  a phenomenon shared by at most one or two other Cepheids
(Bruntt et al.\ 2008; Spreckley \& Stevens 2008).  A variable amplitude
(decreasing and increasing)  is much more likely to be due  to a pulsation
interaction than to an evolutionary effect, which would be expected  to go in
one direction only.  In fact the classical Cepheid with the most prominent 
amplitude variation is V473 Lyr, which is a second-overtone pulsator. 

Neilson et al.\ (2012) have made a population synthesis analysis of period
changes in Cepheids and conclude that both  main sequence core convective
overshoot and mass loss (see below) are needed to match the observations to the
predictions.  This is based both on the size of the period change and also the
proportion of second crossing (period increase)  to third crossing (period
decrease) stars.  However, the working model above  is that many of the short
period stars are overtone pulsators and hence the period changes are determined
by a  pulsation component as well as an evolutionary component. Neilson et al.\ 
draw attention to the fact that short-period stars fit the predictions  less
well than long-period stars.  If the short-period stars  omitted for this
reason, Fig. 3 in Neilson et al.\ shows that for the long-period stars, the
distribution of period changes is well matched by evolutionary predictions.   It
is  likely that this also changes the proportion of positive to negative changes
as well,  but a firm statement requires the detailed identification of overtone
pulsators.   Note further, that the period changes discussed are biased in the
sense that they include only stars with measurable period changes.  A large
fraction of Cepheids  in the Szabados catalogs (1977, 1980, 1981, 1989, 1991)   
show no period change, at least over the 60 years for which we have photometric
observations. Presumably their periods also change as they evolve, but more
slowly.  This means that Figs.~2 and 4 in Neilson et al.\ where the predictions
are larger than the observed values are even more difficult to reconcile with
the expected {\it mean} period changes.  

  The period variation  in Polaris challenges the current models.  
In a number of stars   period
change is not monotonic, but rather a combination of increase and decrease
(e.g., Fig.~3 in Berdnikov et al.\ 1997).  
The simplest explanation for Polaris and other overtone pulsators would be 
that the current epoch reflects, at least partly,
short term unsteadiness rather than long term evolution.  
Some possible scenarios where pulsation might 
influence the perceived period change in overtone pulsators 
 are sketched in Evans et al.\ (2002).

That said, Polaris does have an unusually large period change.  


\subsubsection{Which Instability Strip Crossing?}

Neilson (2014) provides a recent thorough discussion of the observations of
Polaris, together with  evolutionary models, mass loss, rotation, and
abundance.    A question related to the period change is the crossing of the
instability strip.   T13 have asserted that their distance is
consistent with a first crossing.  This, however, has been refuted by van
Leeuwen (2013) and Neilson  (2014) and is not  consistent with our mass summary
above (Table 3).  Following our ``working model''  above that period changes in
overtone pulsators are not entirely due to  simple monotonic evolution  removes
rapid period change as an argument for Polaris as a star  on the first crossing
of the instability strip. 





Anderson (2018) has recently proposed a different interpretation for Polaris, 
that it is in the first crossing of the instability strip and pulsating
in the first overtone.  This is based in the comparison of the  
the luminosity of B18 with evolutionary tracks from the Geneva group
(Anderson et al.\ 2014), as well as CNO abundance enhancements (discussed
below).  A major implication of this interpretation is that the 
mass of the Cepheid is 7 $M_\odot$.  However, mass of Polaris 
based on the new {\it Gaia\/} distance  (3.45 $\pm$ 0.75 $M_\odot$; Table 3)
is inconsistent with this.
  
Thus, based on the period and luminosity, Polaris is on the third crossing of
the instability strip.


\subsubsection{Mass Loss}

Could the current mass of Polaris have been affected by mass loss?
The question of whether Cepheids lose mass, and in particular whether pulsation 
fosters mass loss is an important one.  This has been a suggestion to solve the ``Cepheid
mass problem'', reconciling the larger evolutionary track masses with smaller 
pulsation predictions.  Neilson et al. (2012) and references therein discuss 
studies which indicate that pulsation can drive mass loss.  They 
conclude that mass loss is necessary to 
explain period change predictions (but we argue above for complexities in the 
interpretation of the period change).  
Observationally, there have been a series of interferometric observations 
indicating circumstellar envelopes (CSEs) in the {\it K}-band  (Kervella et al.\ 2006;
M\'erand et al.\ 2006,
2007). This feature seems to be shared by all the Cepheids observed, but not the 
nonvariable yellow supergiant $\alpha$ Per.  These shells are particularly important 
to characterize and understand since they are dependent both on wavelength and 
pulsation period.  Thus they could affect the infrared (IR) Leavitt (period-luminosity)
law and may indicate  mass loss. 
On the other hand, {\it Spitzer} observations have found only occasional indications of 
extended IR emission (Barmby et al.\ 2011).  Specifically, 10\% have fairly
clear emission, with a further 14\% with tentative emission.     
This agrees with the summary of mid-infrared excess by Schmidt (2015).
He finds that only
9\% of classical Cepheids have 
IR excess (omitting 3 with anomalous energy distributions)
and only  stars with periods 11$^d$ or longer. 
Thus, while there is evidence that a small fraction of Cepheids 
have extended IR emission, it is by no means universal.   
Uniquely in the case of  $\delta$~Cep there is also a more extended 
feature in the {\it Spitzer\/} study  
which looks like a bow shock  (Marengo et al.\
2010).  This was followed up by Matthews et al.\ (2012) with a \ion{H}{1} 21~cm 
observation consistent with an outflow.   

For Polaris itself, a CSE has been detected (M\'erand et al.\ 2006).  However, 
 {\it Sptizer} IRAC and MIPS observations have found only a suggestion of
 IR excess in Polaris (Marengo et al.\ 2010) and no 
extended IR emission (Barmby et al.\ 2011). 

Recent X-ray observations have revealed another process which may contribute to  
 upper atmosphere mass motions.   
The passage of the pulsation wave at minimum radius through the photosphere and chromosphere 
is marked disturbances such ultraviolet emission lines, 
most clearly shown in $\delta$~Cep itself (Engle et al.\ 2017).  
X-rays, on the other hand remain low and constant 
at minimum radius, but exhibit a sharp rise and fall just after maximum radius.  One 
possible mechanism for this is a flare such as frequently found on the sun and other
stars with convective surfaces.   
Often flares are associated with coronal mass
ejections.  The burst of X-rays in Cepheids appears to be tied to the pulsation
cycle, above the photosphere, and triggered by the collapse of the atmosphere after
maximum radius.  Thus it is possible that the X-ray bursts may be part of a chain 
that moves material upward in Cepheid atmospheres.  

Polaris is a very low-amplitude Cepheid, 
and as yet an X-ray burst has not been detected (Evans et al.\ 2018).
However, the phase range where it would be expected has not been fully covered by 
X-ray observations.   

\subsubsection{Mergers}

Figure~9 shows the positions of V1334\,Cyg and Polaris\,Aa and B in the
color-magnitude diagram (CMD; reddening-corrected absolute $V$ magnitude versus
$B-V$ color). B18 pointed out that Polaris\,B lies above the main sequence in
the CMD, if the relatively small parallax that they found is assumed. This
remains true even if we use the larger \Gaia\/ parallax, as shown in Figure~9.
Also plotted are MIST isochrones for solar metallicity and zero rotation, for
ages of 100~Myr and 2.25~Gyr. The nominal 100~Myr isochrone fits the locations
of the two Cepheids reasonably well. The 2.25~Gyr isochrone was chosen because
it fits the position of Polaris\,B\null. 
 We have argued in Section 4.2.3 that evolutionary tracks for the ``blue loop'' phase require
some adjustment,  specifically that the observed masses of the two
Cepheids are significantly smaller than predicted by the luminosity
of the isochrones. However, the 100 Myr isochrone in Figure~9 
provides a schematic description of
the evolution of a Cepheid which is  clearly a reasonably massive evolved
star with an age of order 100 Myr. 

The apparent discordance between the ages of Polaris and its wide companion was
discussed by B18. Setting aside the remote possibility that B is not actually a
physical companion of the Cepheid, B18 considered two scenarios: (1)~The Polaris
system is young, and the B component is abnormally luminous. This could
conceivably be due to B being an unresolved binary, even at \HST\/ resolution,
although there is no direct evidence for this. We are currently planning 
an observation with the  CHARA array
to test whether B can be resolved. This might, of course, affect
the {\it Gaia} parallax.  
Or possibly B is temporarily
over-luminous due to a recent stellar merger. (2)~Alternatively, the system may
in fact be old, and it is Polaris itself that appears anomalously young. For
example, Polaris might be a merger product. It certainly seems likely that at
least some Cepheids are descendants of merged binaries, given the high frequency
of close binaries among their B-type progenitors. In fact, Neilson et al.\
(2015) have pointed out an age discrepancy for the binary eclipsing LMC Cepheid
OGLE-LMC-CEP1812, and presented arguments that this Cepheid is a post-merger
product. Based on the MIST 2.25~Gyr isochrone, the mass of Polaris\,B is
$\sim$$1.4\,M_\odot$, and stars of $\sim$$1.6\,M_\odot$ have essentially
completed their evolution. Since we find a dynamical mass for Polaris\,Aa of
$3.45\pm0.75\,M_\odot$, a merger scenario based on two stars both near the
main-sequence turnoff at $\sim$$1.5\,M_\odot$ appears possible. 

We investigated whether Polaris\,Ab provides any useful constraint on these two
scenarios. For example, if Ab also appeared to be an old star, that would
greatly strengthen the merger scenario for the Cepheid progenitor.
Unfortunately, other than the dynamical mass of $1.63\pm0.49\,M_\odot$, the only
information we have on the nature of Polaris\,Ab is its absolute magnitude in
the F220W (2200~\AA) bandpass, which we measure in our ACS/HRC frames to be
about +4.25 (Vega scale, \Gaia\/ distance). We have no direct information on its
color or spectral type. The MIST isochrones provide absolute magnitudes in the
F220W bandpass, but such an absolute magnitude can be attained by a star of this
mass with either the young or old isochrones plotted in Figure~9.

\subsubsection{Abundances}

Abundances provide clues to the history of an evolved star.  Specifically, 
an increase in N and a decrease in C compared with main sequence 
values indicates processed material.  The abundances of Polaris have 
been measured several times, summarized by Usenko et al.\ (2005), 
including Luck \& Bond (1986).  Similarly, abundances for V1334 Cyg 
have been measured by Takeda et al.\ (2013).  Abundances from Usenko et al.\
and Takeda et al.\ are summarized in Table 4, showing very similar
amounts N enrichment and C decrease suggesting comparable processing 
mechanisms and quantity.   
As discussed by Neilson (2014), the 
 processing of material could come 
either through rotational mixing near the main sequence or dredge-up 
as a star becomes a red giant.  He computes evolutionary tracks with 
a range of main sequence rotational velocities. From these predictions
and the relatively low observed rotational velocity he  concludes that
rotational mixing is not responsible for the abundance distribution
hence the star must be beyond the first crossing of the instability strip.  
Anderson (2018) on the other hand comes to the opposite conclusion 
from the abundances, that they are consistent with a first crossing.
We note that Figs. 8, 10, and 11 in Anderson et al.\ (2014) show that
the N and C abundances are roughly in agreement with either pre-dredge-up
with a lot of initial rotation or post-dredge-up with little initial rotation.  
 Takeda et al.\ surveyed 12 Cepheids at several 
phases in the pulsation cycle for each.  The C and N abundances  
(in the mean) also agree with those of V1334 Cyg and Polaris, providing
no evidence for an evolutionary state other than the post-dredge-up 
state typical of Cepheids.


A further mechanism which would produce processed material is the merger of a 
close main sequence binary (previous section).


\subsection{Future Work}

These results conclude the work possible with \HST\/ in resolving 
Polaris Aa and Ab
since the separation between the 
two stars becomes progressively smaller during the following  years. 
In this discussion we are 
providing  a summary of the understanding of the system at present.  Clearly 
continued study of the orbit over the next few years will provide an improved 
definition of the orbit and the mass. These observations are to be continued with the 
interferometry (e.g., the 
 CHARA array).  In fact a tentative detection has been made, which is in reasonable agreement
with the current orbit.

\section{Summary}


To summarize, Polaris now has a  suite of directly measured parameters to identify its
physical state: luminosity, mass, radius, period change, and abundance, although 
we stress that the mass presented here is preliminary, and Tables 2 and 3 
illustrate the dependence on the data string and the distance.  The mass is within
approximately 1$\sigma$ of the mass of V1334 Cyg, which is more precisely measured.
However, 
for both Polaris and V1334 Cyg, the masses are smaller than predicted by 
evolutionary tracks for the observed  luminosities.
   There are several suggestions
of the more complex history for Polaris in arriving at its present configuration. 
It has a circumstellar envelope in the infrared, making it likely that 
some mass has been lost.  The interpretation of the rapid period 
change is likely to involve more than  evolution through the 
instability strip alone, but also may be characteristic of 
overtone pulsation.  In agreement with this is the fact that V1334 Cyg
(also an overtone pulsator) has no measurable period change.  The difference in 
age required for  isochrones 
for Polaris Aa and B suggests  that the Cepheid 
might be a merger product.  Probing these questions is the goal of 
further observations of the orbit.

\acknowledgments

It is a pleasure to thank Pierre Kervella for valuable comments 
on the draft.  
Support for this work was provided by NASA
through  grants from the Space Telescope Science Institute
(HST-GO-10891.01-A and
HST-GO-13369.001-A), which is
operated by the Association of Universities for Research in Astronomy,
Incorporated, under NASA contract NAS5-26555.
Support for this work was also provided  from the Chandra X-ray Center NASA 
Contract NAS8-03060.  G.H.S. acknowledges support from NSF Grant AST-1411654.
Vizier and SIMBAD were used in the preparation of this study.
This work has made use of data from the European Space Agency (ESA) mission
{\it Gaia} (\url{https://www.cosmos.esa.int/gaia}), processed by the {\it Gaia}
Data Processing and Analysis Consortium (DPAC,
\url{https://www.cosmos.esa.int/web/gaia/dpac/consortium}). Funding for the DPAC
has been provided by national institutions, in particular the institutions
participating in the {\it Gaia} Multilateral Agreement.

\begin{figure}
\begin{center}
\fbox{\includegraphics[width=3.5in]{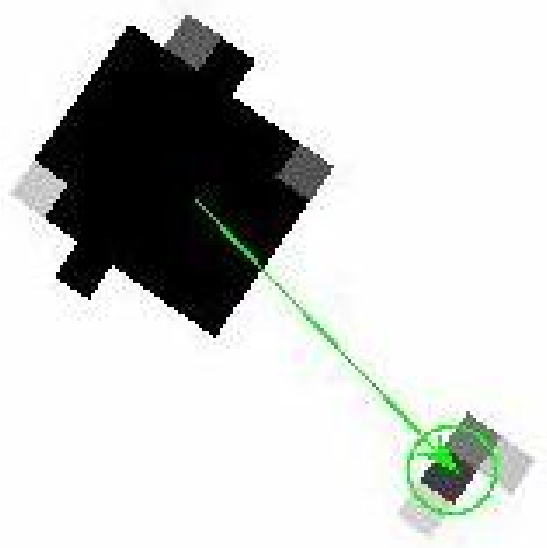}}
\caption{ 
The 2007 WFPC2 image of Polaris, AstroDrizzled to a scale of $0\farcs02\,\rm
pixel^{-1}$, and deconvolved using an image of Polaris~B as the PSF reference.
The faint companion Polaris Ab is detected at a separation  of $0\farcs18$. The
green circle marking the companion has a radius of $0\farcs02$ and the green
arrow indicates the separation of Polaris~Ab from the center of the Cepheid
Polaris~Aa. In Figures~1 through 3, north is at the top and east at the
left.\label{de.07}
}
\end{center}
\end{figure}

\begin{figure}
\begin{center}
\fbox{\includegraphics[width=3.5in]{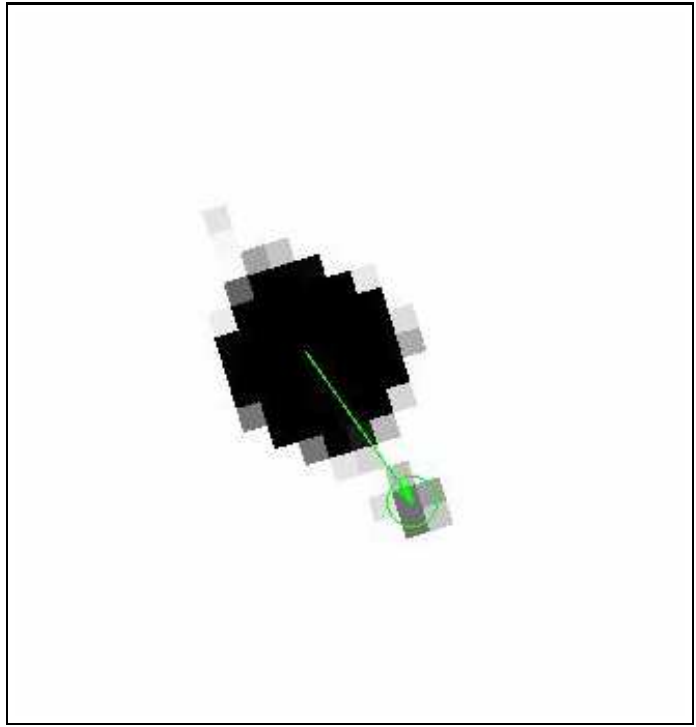}}
\caption{ 
The 2009 WFC3 image of Polaris, AstroDrizzled to a scale of $0\farcs02\,\rm
pixel^{-1}$, and again deconvolved using an image of Polaris~B as the PSF
reference. The companion is detected at a separation of $0\farcs15$. The green
circle marking the companion again has a radius of $0\farcs02$ and the green
arrow indicates the separation of Polaris~Ab from Polaris~Aa. Note slight
elongation of the image due to ``shutter jitter'' in the short WFC3 exposures.
\label{de.09}
}
\end{center}
\end{figure}

\begin{figure}
\begin{center}
\fbox{\includegraphics[width=3.5in]{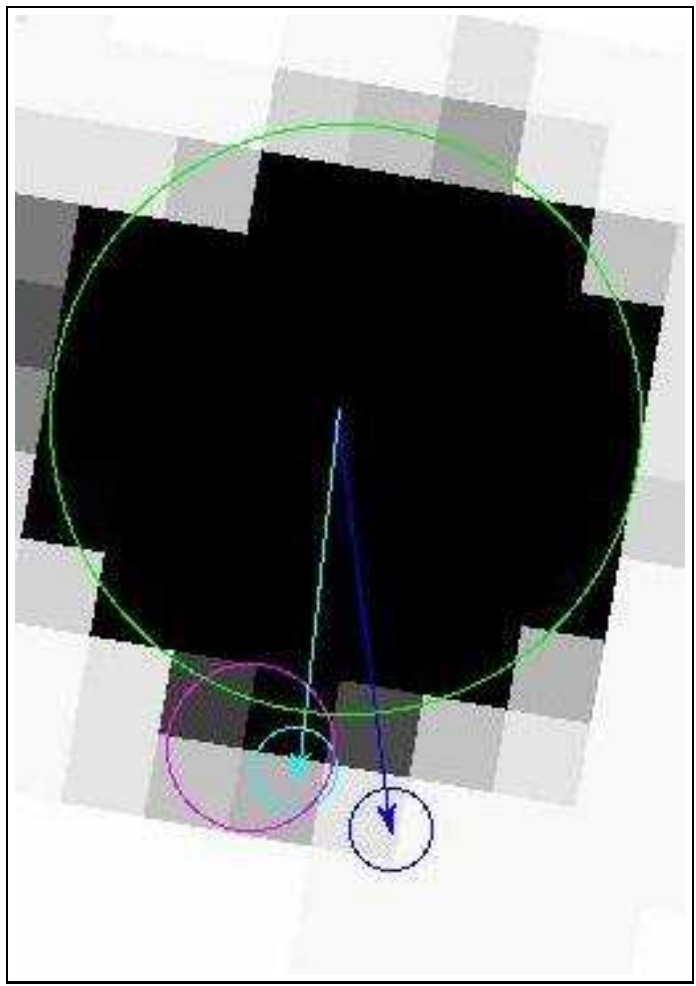}}
\caption{
The 2014 WFC3 image of Polaris, AstroDrizzled to a scale of $0\farcs02\,\rm
pixel^{-1}$. This image has been deconvolved using the EMC2 algorithm (see
text), with images of $\gamma$~Cyg serving as the PSF reference. Note the
magnified scale of this image compared to Figures 1 and 2. Here the large green
circle is to guide the eye to show the  location of the Cepheid. The magenta
circle is where we believe we have  detected the companion, lying $0\farcs08$
from the Cepheid. The dark blue circle and arrow are the predicted location
based on a combination of the spectroscopic orbit, our \HST\/ observations from
2005 to 2007, and the astrometric measurements from Wielen et al.\ (2000). This
location is clearly offset from the brightest  pixels in the area where the
companion is detected.  The light blue circle and orbit are the  predictions
from the same data varying the inclination by 1$\sigma$.    
\label{de.14}
}
\end{center}
\end{figure}

\begin{figure}
\begin{center}
\includegraphics[width=4.5in]{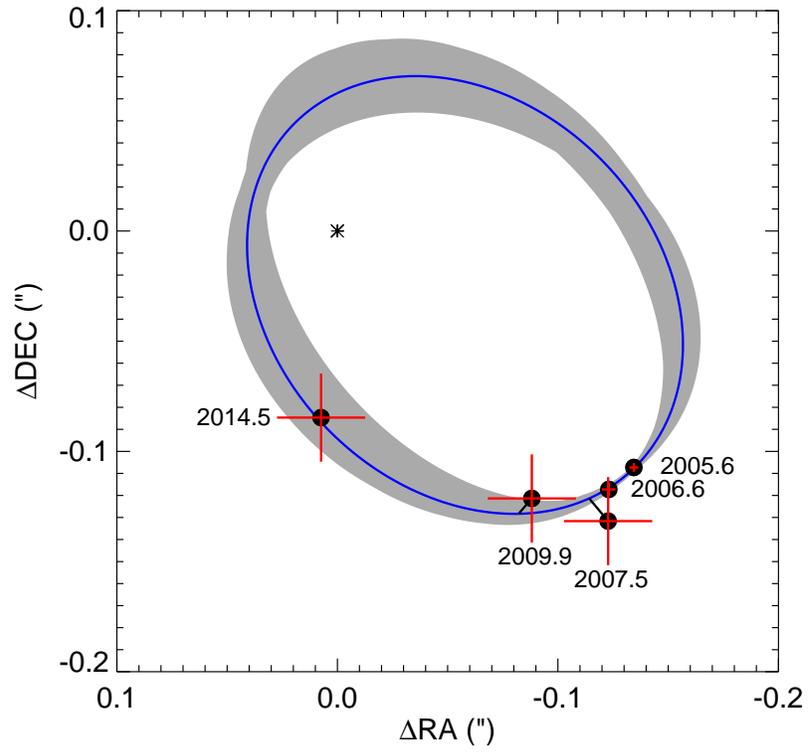}
\caption{
Visual orbit of Polaris Ab relative to Aa, based on our \HST\/ measurements.
Black circles show the measured positions. The sizes and orientations of the
error ellipses are shown as red lines. The blue curve shows the best-fit orbit,
determined as described in the text. The shaded gray area shows the range of
orbital solutions that fit the data. 
}
\label{fig.orbit}
\end{center}
\end{figure}

\begin{figure}
%
\plotone{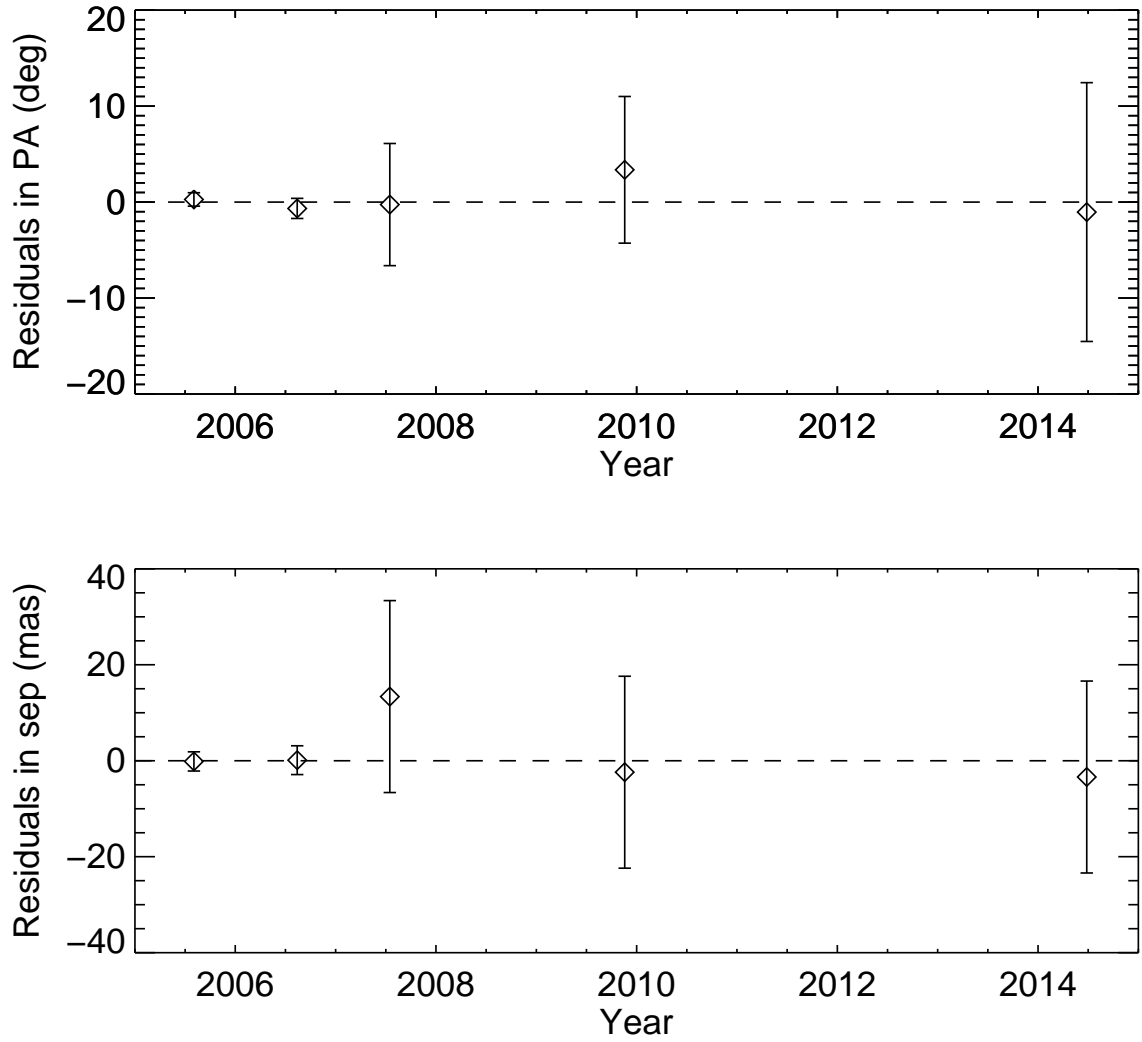}
\caption{Residuals in the position angle and separation measured with {\it
HST} and the best-fit orbital solution.}
\label{fig.residual}
\end{figure}


\begin{figure}
\plotone{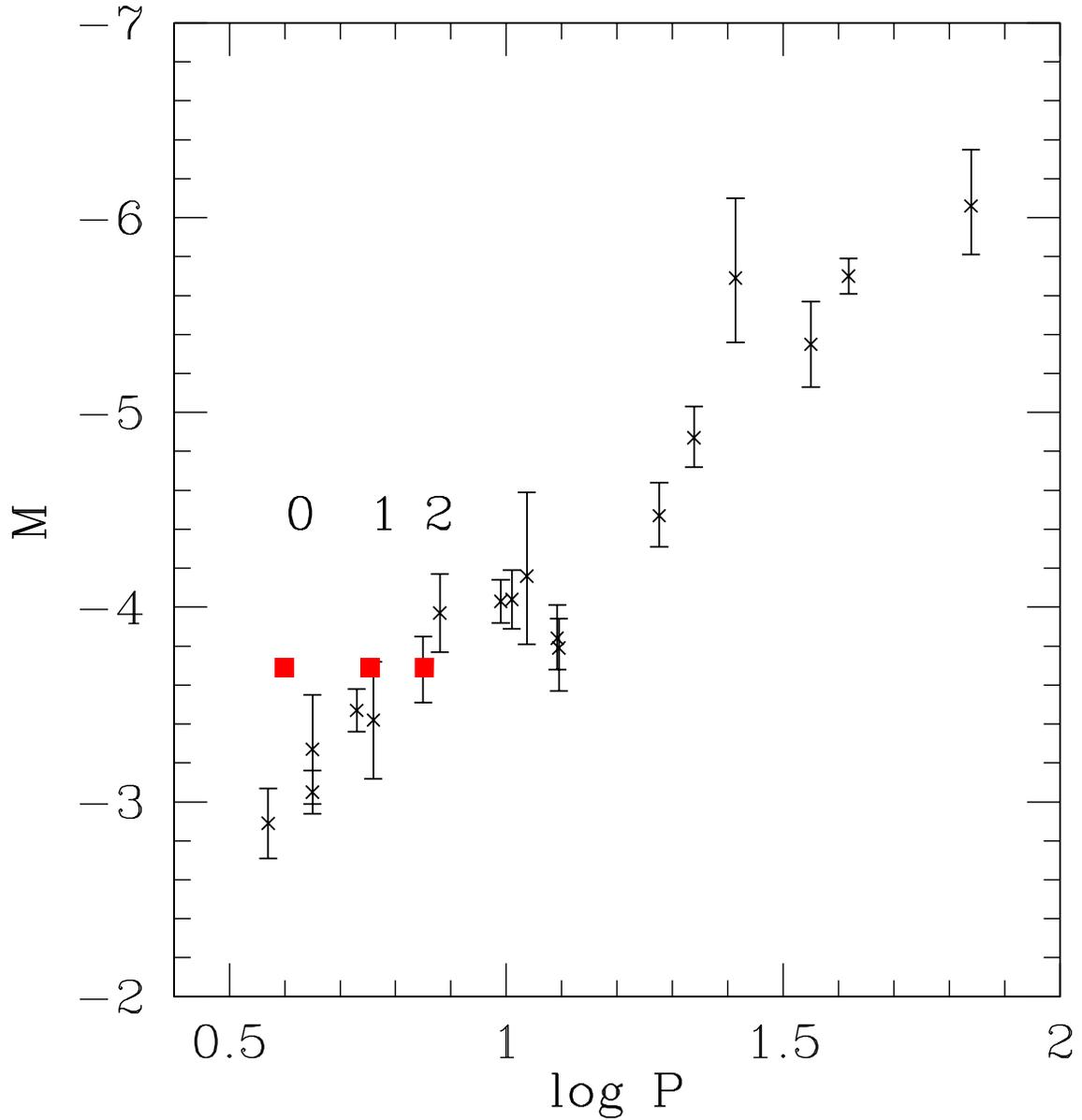}
\caption{Polaris in the context of the Leavitt Law. Polaris  $M_V$ from 
{\it Gaia}: filled red squares;  other Cepheids: x's 
from FGS, {\it HST\/} spatial scans, and RS Pup (see text).
  The labels above the values for Polaris 
indicate the pulsation period adjusted (left to right) to the fundamental period for 
a star pulsating in  the fundamental mode, the first 
overtone, and the second overtone.  Note that the errors on the squares are
smaller than the synbols (including one overlapping an x giving the appearance
of a large error). M$_V$ is in magnitudes; P is in days.
  \label{pol.mode}}
\end{figure}

\begin{figure}
\plotone{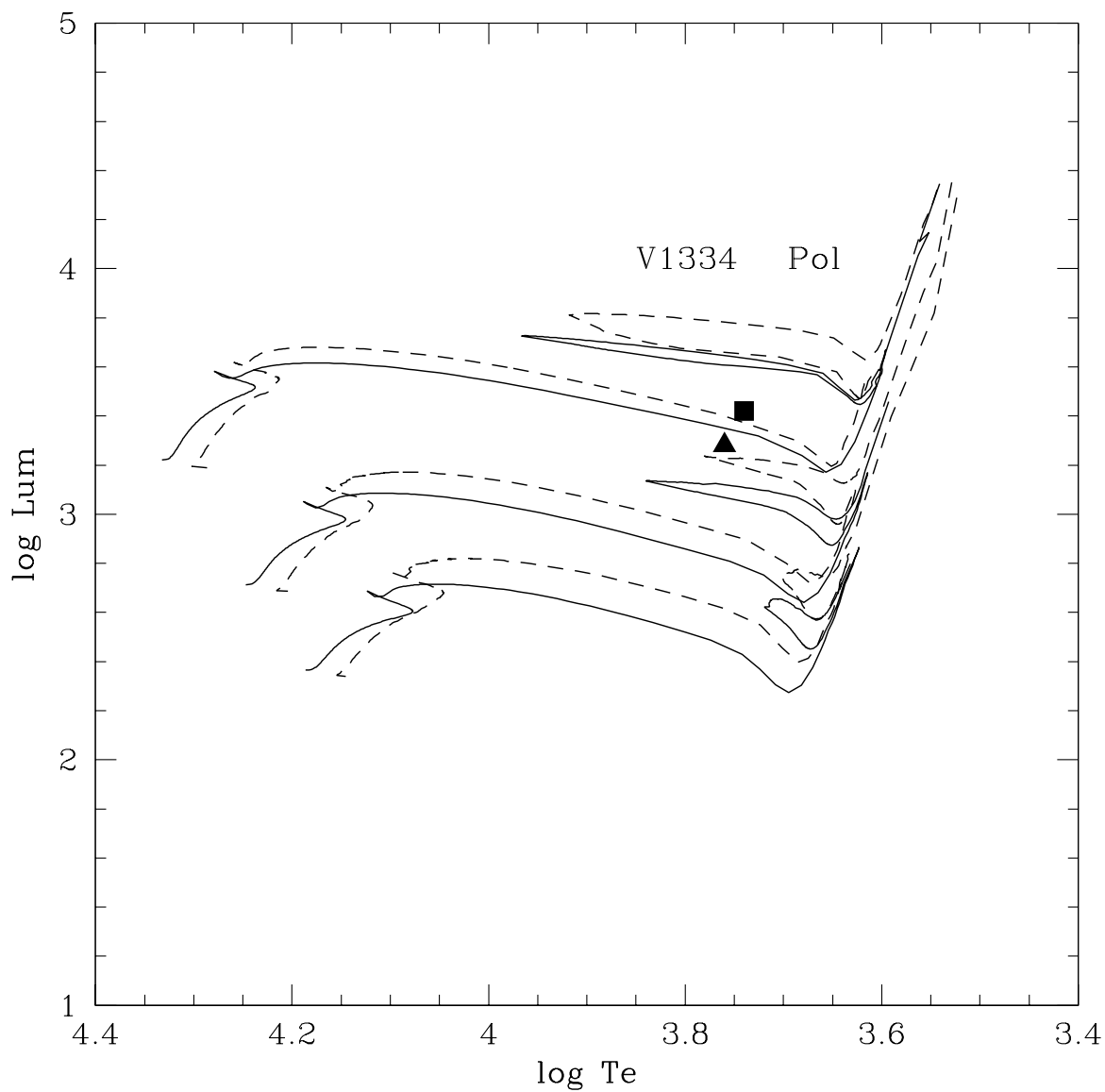}
\caption{Polaris and V1334 Cyg compared with evolutionary tracks.
For Polaris (filled square) and V1334 Cyg (filled triangle) the errors are 
error smaller than the symbols.  The relative positions of Polaris and 
V1334 Cyg are shown by the labels.
The evolutionary tracks from Georgy et al.
(2013) are 4, 5, and 7 $M_\odot$ from bottom to top. Solid lines are for 0 rotational
velocity on the main sequence; dashed lines have 0.95 breakup velocity.
Luminosity is in solar units; temperature is in K.  
  \label{ev.pol}}
\end{figure}

\begin{figure}
\plotone{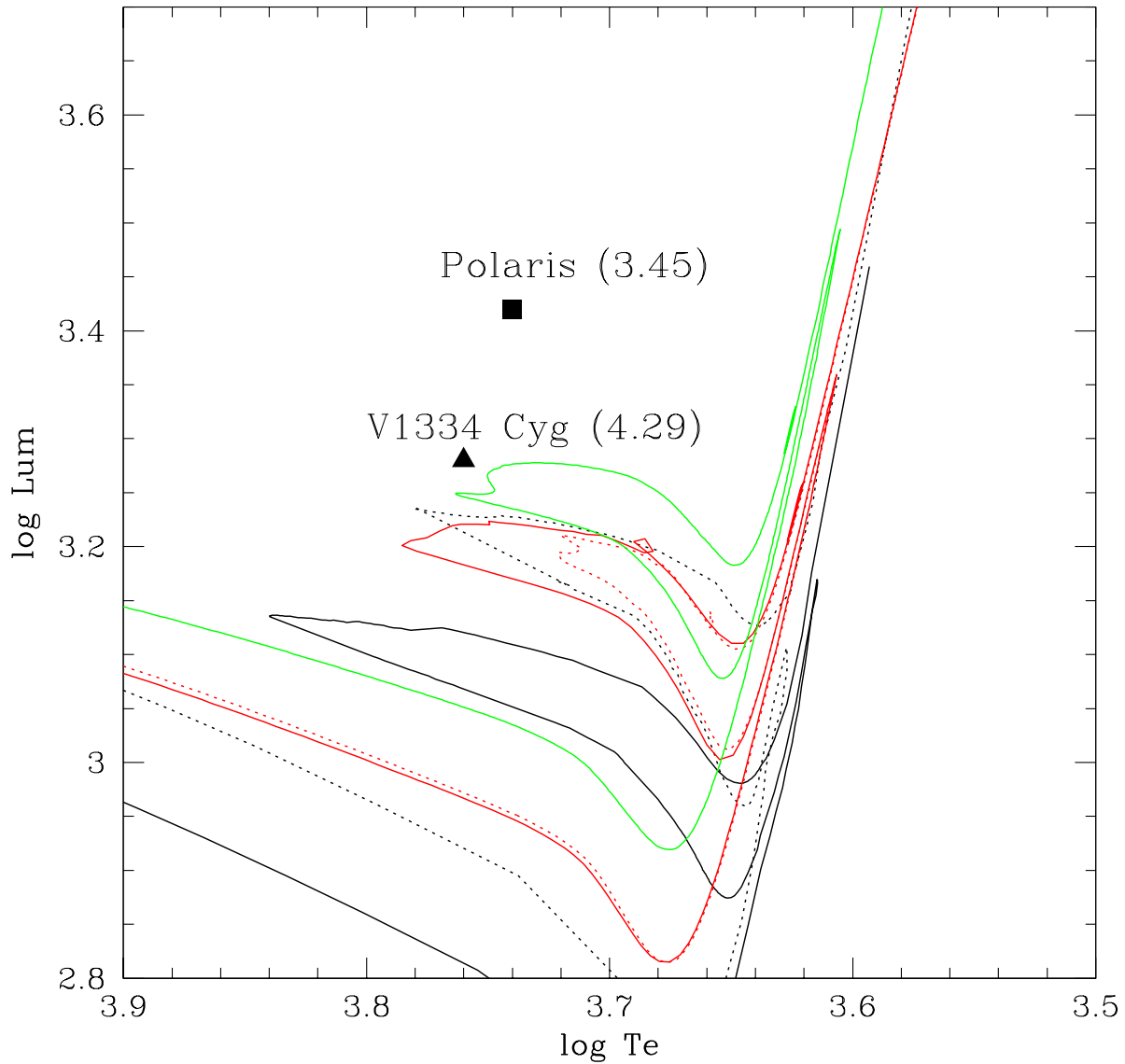}
\caption{Polaris and V1334 Cyg compared with evolutionary tracks 
in the region of the ``blue loops''.
For Polaris (filled square) and V1334 Cyg (filled triangle) the errors are 
smaller than the symbols. The mass in $M_\odot$ is listed next to the star 
name. The evolutionary tracks from Georgy et al.\
(2013): black;  MIST: red; and PARSEC (green), all for  $5\,M_\odot$.  Solid 
and dotted lines represent no rotation and significant rotation respectively
(see text).
Luminosity is in solar units; temperature is in K.  
  \label{ev.pol.blu}}
\end{figure}

\begin{figure}
\begin{center}
\includegraphics[height=5.25in]{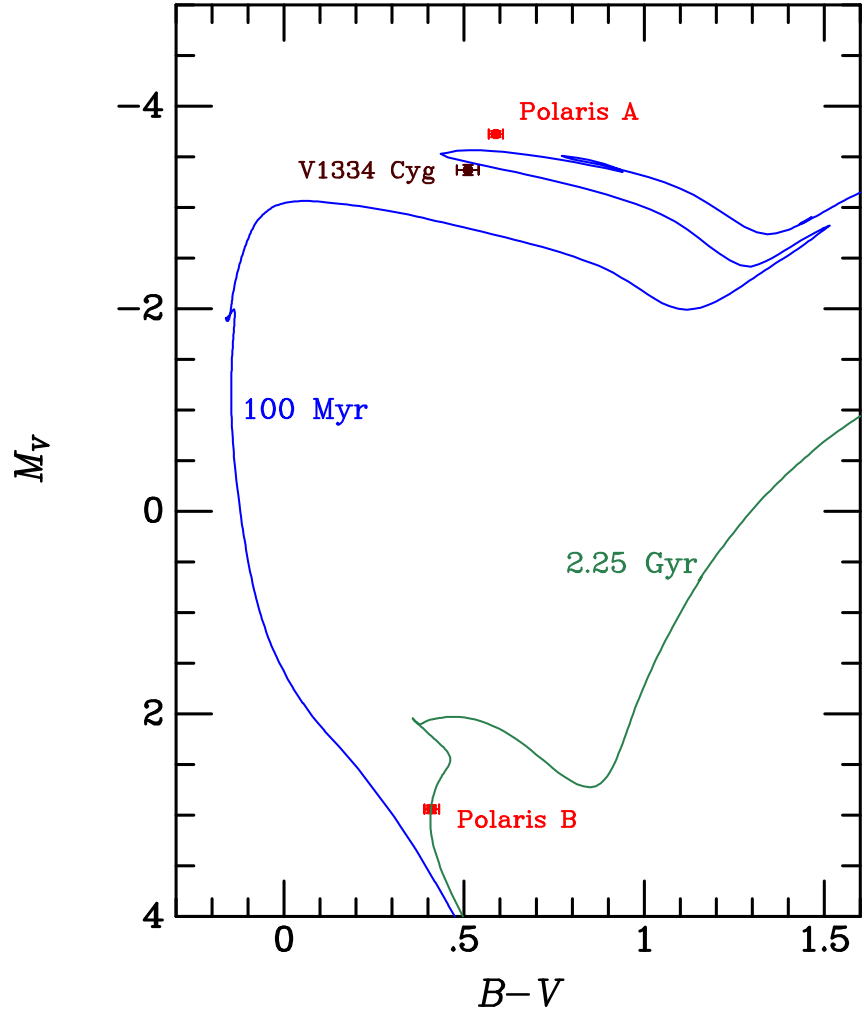}
\end{center}
\caption{
Locations in the color-magnitude diagram (absolute $V$ magnitude vs.\
$B-V$ color) of the Cepheid V1334\,Cyg (purple point) and Polaris\,Aa and B (red
points). Also shown are solar-metallicity and zero-rotation isochrones from the
MIST data base (see text) for ages of 100 Myr (blue line) and 2.25~Gyr (green
line).   
  \label{iso.pol}}
\end{figure}

\clearpage

\begin{deluxetable}{lcccccc}
\tabletypesize{\footnotesize}
\tablecaption{\HST\/ Imaging Observations and Measurements of Polaris Aa--Ab\label{obs}}
\tablewidth{0pt}
\tablehead{
\colhead{UT Date} & \colhead{Besselian} & \colhead{Program}  & \colhead{Instrument} & \colhead{Filter} &
\colhead{P.A. (J2000)\tablenotemark{b}} & \colhead{Separation}     \\
\colhead{} & \colhead{Date} & \colhead{ID\tablenotemark{a}} & \colhead{} & \colhead{} & \colhead{[degrees]} &
  \colhead{[arcsec]}   \\
}
\startdata
2005 Aug  2--3\tablenotemark{c} & 2005.5880 & GO-10593 & ACS/HRC  & F220W  & 231.4 $\pm$ 0.7 &  0.172 $\pm$ 0.002 \\
2006 Aug 13\tablenotemark{c} & 2006.6172 & GO-10891 & ACS/HRC  & F220W  & 226.4 $\pm$ 1.0 &  0.170 $\pm$ 0.003 \\
2007 Jul 17\tablenotemark{d} & 2007.5402 & GO-11293 & WFPC2/PC & F218W  & 223   $\pm$ 6.4   &  0.18 $\pm$ 0.02  \\
2009 Nov 18\tablenotemark{d} & 2009.8813 & GO-11783 & WFC3/UVIS     & FQ232N & 216   $\pm$ 7.6   &  0.15 $\pm$ 0.02  \\
2014 Jun 26\tablenotemark{e} & 2014.4856 & GO-13369 & WFC3/UVIS     & FQ232N & 175   $\pm$ 13.5   &  0.085 $\pm$ 0.02  \\
\enddata
\tablenotetext{a}{N.R.E. was Principal Investigator for all of these programs.}
\tablenotetext{b}{Note that the P.A. is referred to the J2000 equator, not to
the equator of the observation epoch as is the usual custom in ground-based
visual-binary work.}
\tablenotetext{c}{Information for the first two observations is quoted from
Evans et al.\ 2008.}
\tablenotetext{d}{In 2007 and 2009 we also obtained images of
Polaris~B, for use as a PSF reference.}
\tablenotetext{e}{WFC3 observations were also obtained on 2014 Mar 17, but were
not used in the analysis (see text). The 2014 observations were accompanied by a
set of nearly identical exposures on $\gamma$~Cyg for PSF reference purposes.}
\end{deluxetable}


\begin{deluxetable}{lccl}
\tablecaption{Orbital Parameters for Polaris Aa--Ab} 
\tablewidth{0pt}
\tablehead{
\colhead{Parameter} & \colhead{Value} & \colhead{Value} & \colhead{Reference} \\
\colhead{}   & \colhead{(All data)}   & \colhead{(Omitting 2014)}  & \colhead{} \\
}
\startdata
$P$ [yr]             & 29.59 $\pm$ 0.02   & 29.59 $\pm$ 0.02   & Kamper (1996) \\
$T_0$                & 1987.66 $\pm$ 0.13  & 1987.66 $\pm$ 0.13 & Wielen et al.\ (2000) \\
$e$                  & 0.608 $\pm$ 0.005  & 0.608 $\pm$ 0.005  & Kamper (1996) \\
$\omega_{Aa}$ $[^\circ]$\tablenotemark{a} & 303.01 $\pm$ 0.75  & 303.01 $\pm$ 0.75  & Kamper (1996) \\
$a$ $['']$           &  0.1204  $\pm$   0.0059  &  0.1215  $\pm$  0.0062   & This work \\
$i$ $[^\circ]$     &   146.2  $\pm$    10.9  &  144.1   $\pm$  10.0  & This work \\
$\Omega$ $[^\circ]$\tablenotemark{b}   &  191.4  $\pm$    4.9   &    192.1  $\pm$   4.7  & This work \\
$\chi^2_\nu$         &       0.179           & 0.241  & This work \\
\enddata
\tablenotetext{a}{The angle between the ascending node and periastron passage, as referenced to Polaris Ab is given by $\omega_{Ab} = \omega_{Aa} + 180^\circ = 123.01^\circ$.}
\tablenotetext{b}{$\Omega$ is flipped by 180 deg compared with the value reported in Table 4 
of Evans et al. (2008) because the earlier paper defined $\omega_{Ab}$ = 303.01 deg in the visual 
orbit fit, rather than using the standard spectroscopic convention of referencing $\omega$ to 
the primary.}
\end{deluxetable}

\begin{deluxetable}{lcc}
\tablecaption{Dynamical Masses for Polaris Aa and Ab for Different Adopted
Parallaxes} 
\tablewidth{0pt}
\tablehead{
%
\colhead{Parameter} & \multicolumn{2}{c}{--- Parallax adopted from:\tablenotemark{a} ---} \\
\colhead{         }  & \colhead{vL07}  & \colhead{G18} \\
}
\startdata
$\pi$ [mas]              &  $7.54\pm0.11$  &  $ 7.3218\pm0.0281 $  \\
$d$ [pc]                   & $ 132.6\pm1.9 $  & $ 136.58\pm0.52 $   \\
$M_{\rm tot}$ [$M_\odot$]  & $4.65\pm 0.71$   & $ 5.07\pm 0.75 $ \\
$M_{\rm Aa}$ [$M_\odot$]   & $ 3.11\pm0.70 $  & $ 3.45\pm 0.75 $ \\
$M_{\rm Ab}$ [$M_\odot$]  &  $1.54\pm 0.46 $  &  $1.63\pm0.49 $ \\
$\langle M_{V,\rm Aa} \rangle$\tablenotemark{b}  & $-3.66\pm0.04$  &
$-3.73\pm0.03  $  \\
\enddata
\tablenotetext{a}{Abbreviations for sources of adopted parallaxes are 
 vL07 = van Leeuwen 2007; 
G18 = {\it Gaia} including Lindegren correction.}
\tablenotetext{b}{Mean visual absolute magnitude of Polaris~Aa, assuming
$\langle V\rangle = 1.982$, $E(B-V)=0.01\pm0.01$, and the adopted parallax.} 
\end{deluxetable}

\begin{deluxetable}{lccc}
\tablecaption{Abundances} 
\tablewidth{0pt}
\tablehead{
\colhead{         } & \colhead{[C/H]} & \colhead{[N/H]} & \colhead{[O/H]} \\
}
\startdata
Polaris Aa & $-0.17$	      &  +0.42  	  & $-0.00$	     \\
V1334 Cyg  & $-0.24 \pm 0.03$ & $+0.44 \pm 0.03$  & $+0.03 \pm 0.03$ \\
\enddata
\end{deluxetable}

\end{document}